\documentclass[12pt]{amsart}
\usepackage{epsf}
\usepackage{psfrag}
\usepackage{fullpage}
\usepackage{float}
\usepackage{mathrsfs}
\usepackage{amsfonts}
\usepackage[centertags]{amsmath}
\usepackage{amssymb}
\usepackage{amsthm}
\usepackage{graphicx}
\usepackage{float}
\usepackage[all]{xy}
\usepackage{tikz-cd}
\usepackage{tikz}
\usetikzlibrary[shapes]
\usepackage{multirow}
\usepackage{caption}
\usepackage{xparse}
\usepackage{lscape}
\usepackage{dsfont}
\usepackage{inputenc}
\usepackage{subcaption} 
\usepackage{enumerate,enumitem}
\usepackage[margin=1in]{geometry} 
\usetikzlibrary{matrix,arrows,decorations.pathmorphing}
\usepackage{listing} 
\usepackage{hyperref}
\hypersetup{
bookmarksnumbered,
pdfstartview={FitH},
breaklinks=true,
linkcolor=blue,
urlcolor=blue,
citecolor=blue,
bookmarksdepth=2
}
\usepackage{cleveref}
\usepackage{ytableau}
\usepackage{adjustbox}
\usepackage[normalem]{ulem}
\usepackage{cancel}

\theoremstyle{plain}
   
   \newtheorem{theorem}{Theorem}[section]

   \newtheorem{corollary}[theorem]{Corollary}

\theoremstyle{definition}
   \newtheorem{definition}[theorem]{Definition}
   
   \newtheorem{example}[theorem]{Example}

   \newtheorem{remark}[theorem]{Remark}

\numberwithin{equation}{section}

\newcommand{\CC}{{\mathbb {C}}}

\newcommand{\ZZ}{{\mathbb {Z}}}

\newcommand{\ch}{{\operatorname{ch}}}

\newcommand{\SSYT}{{\operatorname{SSYT}}}

\DeclareMathOperator{\Gr}{Gr} 
\DeclareMathOperator{\GL}{GL}

\begin{document}

\title{Dual conformal invariant kinematics and folding of Grassmannian cluster algebras}

\author{Jian-Rong Li, Changjian Su, and Qinglin Yang}
\address{Jian-Rong Li, Faculty of Mathematics, University of Vienna, Oskar-Morgenstern Platz 1, 1090 Vienna, Austria}
\email{lijr07@gmail.com}
\address{Changjian Su, Yau Mathematical Sciences Center, Tsinghua University, Beijing, China}
\email{changjiansu@mail.tsinghua.edu.cn}
\address{Qinglin Yang, Max–Planck–Institut f\"ur Physik, Werner–Heisenberg–Institut, D–85748 Garching bei M\"unchen, Germany}
\email{qlyang@mpp.mpg.de}

\date{}

\begin{abstract}
Grassmannian manifolds $\Gr(4,n)$ are closely related to the kinematic
space of $n$-particle scattering processes in $D=4$, and their
combinatorial and geometric structures have played an important role in
the study of conformal invariant theories and scattering amplitudes.
He, Li, and Yang \cite{HLY26} observed that restricting $D=4$ kinematics
to a $D=3$ subspace can be interpreted as a folding of the Grassmannian
cluster algebra $\CC[\Gr(4,n)]$ for $n\leq 8$. In this paper, we derive
general expressions for the $D=3$ kinematic constraints in terms of
Pl\"ucker coordinates of $\Gr(4,n)$ directly from the three-dimensional
kinematic condition. We then construct a family of foldable seeds for
$\CC[\Gr(4,n)]$, obtained explicitly from the standard initial
seed by mutation, whose folding conditions reproduce these
kinematic constraints. This establishes the connection between $D=3$ kinematics
and folding of Grassmannian cluster algebras for general $n$.
\end{abstract}

\maketitle

\section{Introduction}

Nowadays, physicists are gaining knowledge about particle physics at high
energies by studying scattering processes experimentally at colliders such as
the Large Hadron Collider (LHC), while also developing theoretical methods to
predict the outcomes of such experiments. Scattering amplitudes play a crucial
role in this program, as they encode the basic information about particle
scattering processes and provide a bridge between experiment and theory.
By studying scattering amplitudes in different quantum field theories (QFTs),
especially those of quantum chromodynamics (QCD), one can not only test
theoretical models of fundamental particles by comparing analytical
predictions with experimental data, but also uncover new physical and
mathematical structures in QFT and deepen our understanding of fundamental
physics.

Generally speaking, scattering amplitudes
$A_n(g,p_1,\ldots,p_n)$ are complicated but highly structured functions of
the coupling constant $g$, the momenta
$p_1,\ldots,p_n\in\mathbb{C}^D$, and other quantum numbers associated with
the particles participating in the scattering process. Here $D$ denotes the
spacetime dimension of the QFT. Our physical spacetime is four-dimensional,
consisting of three spatial dimensions and one time dimension. Since the
development of QFT, physicists have developed many powerful tools for
studying scattering amplitudes. For example, in perturbation theory, after
expanding in the coupling constant, amplitudes at each loop order can be
expressed in terms of Feynman integrals, whose integrands are rational
functions of external data and internal loop momenta. More advanced methods
for studying scattering amplitudes can be found, for example, in the review
\cite{EH13}.

For physically relevant theories such as QCD, however, obtaining analytic
expressions for scattering amplitudes can be extremely difficult. This
motivates the study of simpler but still highly nontrivial theories as
laboratories in which new methods and structures can be explored before being
applied to more general theories. One particularly important example is
planar $D=4$ maximal supersymmetric Yang--Mills theory with $\mathcal{N}=4$ supersymmetry ($\mathcal{N}=4$ SYM) \cite{ACK08}, whose scattering amplitudes have revealed
a rich collection of physical and mathematical structures
(see, for example, \cite{ADMST22} for a recent review). In particular, the
dual conformal symmetry \cite{DHKS08} of the theory implies that its
$n$-particle scattering amplitudes can be expressed in terms of
dual-conformal-invariant kinematic variables. After introducing momentum
twistors \cite{Hod}, the corresponding kinematic space can be described by $\Gr(4,n)/\GL(1)^{n-1}$,
where the quotient accounts for the projective rescaling of each momentum
twistor \cite{ABCGPT16}. This connection between scattering kinematics and
the Grassmannian has led to rich geometric and cluster-algebraic structures
in scattering amplitudes and Feynman integrals. These Grassmannian and cluster-algebraic structures have inspired extensive studies of scattering amplitudes and Feynman integrals, including the cluster bootstrap program; see, for example, \cite{CDD+}.
In particular, \cite{HLY26}
introduced ``kinematics quivers'' and established relations between
$D=4$ dual-conformal-invariant kinematics and subalgebras of the Grassmannian cluster
algebra $\mathbb{C}[\Gr(4,n)]$ for $n \le 8$. 

Although physical scattering processes take place in four-dimensional
spacetime, special kinematic limits of four-dimensional theories can exhibit
lower-dimensional kinematics. In particular, $D=3$ kinematics can arise by
restricting the external momenta to a three-dimensional Minkowski subspace.
Such a limit is relevant in the study of four-dimensional scattering
amplitudes and Feynman integrals. Three-dimensional scattering theories also
provide an important context in which dual-conformal kinematics arises, such
as the $D=3$ planar $\mathcal{N}=6$ Chern--Simons matter theory
(Aharony--Bergman--Jafferis--Maldacena theory) \cite{ABJM08}. In this paper,
we focus on the former setting, namely the $D=3$ reduction of
dual-conformal-invariant $D=4$ kinematics.

There are several equivalent ways to impose such a $D=3$ reduction on
four-dimensional kinematics. For example, one may set one spatial component
of each momentum to zero, impose Gram determinant constraints, or impose the
corresponding symplectic condition on the momentum-twistor variables
\cite{EHK+}. In \cite{HLY26}, the authors observed that this $D=3$ reduction
can also be described in terms of a folding of the cluster algebra
$\mathbb{C}[\Gr(4,n)]$ for $n \le 8$. 

In this paper, we study the $D=3$ reduction of dual-conformal-invariant
$D=4$ kinematics and its relation to the Grassmannian cluster algebra $\CC[\Gr(4,n)]$ for general $n$. We first derive the corresponding constraints directly from the
$D=3$ restriction of the four-dimensional spinor-helicity variables.
In terms of momentum-twistor coordinates, these constraints take the form
\begin{equation}
\label{folding1}
    \frac{
    P_{a,a+1,a+2,c}P_{a-1,a,a+1,c+1}
    }{
    P_{a-1,a,a+1,a+2}P_{a,a+1,c,c+1}
    }
    =
    \frac{
    P_{a+1,c-1,c,c+1}P_{a,c,c+1,c+2}
    }{
    P_{c-1,c,c+1,c+2}P_{a,a+1,c,c+1}
    } ,
\end{equation}
for $c-a=i+3$, $i\in[0,n-6]$, $a,c\in[n]$,
where we use the notation $[a,b]=\{a,a+1,\ldots,b\}$, $a\leq b$, and $P_{a,b,c,d}$ is a Pl\"{u}cker coordinate, see Section \ref{subsec:grassmannian cluster algebras initial seed}. 

Scott \cite{Sco} proved that the coordinate ring $\mathbb{C}[\Gr(k,n)]$ admits a cluster algebra structure, known as the Grassmannian cluster algebra. One standard initial seed of
$\mathbb{C}[\Gr(k,n)]$ is given by a quiver of rectangular shape consisting
of triangles, with initial cluster variables given by certain Pl\"ucker
coordinates; see, for example, Figure~\ref{fig:initial seed Gr49}. For every
$r\in\mathbb{Z}_{\geq2}$ and $2r\leq n-2$, we construct an explicit mutation
sequence that transforms the initial seed of
$\mathbb{C}[\Gr(2r,n)]$ into a seed whose quiver has a rectangular shape
consisting of squares. Moreover, the mutable part of the resulting quiver
is symmetric. The cluster variables in this seed are again Pl\"ucker
coordinates. The resulting quiver can then be folded, and by identifying
pairs of cluster $X$-coordinates we obtain a collection of equations. In
the case of $\mathbb{C}[\Gr(4,n)]$, these equations are precisely the
constraints \eqref{folding1}. This provides a cluster-algebraic
realization of the $D=3$ reduction of the $D=4$ kinematic space.

The paper is organized as follows. In Section \ref{sec:$D{=}3$ kinematics limit}, we describe the condition which constrains the general $D{=}4$ kinematics in a scattering process to $D{=}3$ subspace. In Section \ref{sec:Grassmannian cluster algebras}, we recall results of Grassmannian cluster algebras. In Section \ref{sec:folding of Grkn}, we show that the condition which constrains the general $D{=}4$ kinematics in a scattering process to $D{=}3$ subspace could be understood as folding of Grassmannian cluster algebras. 

\subsection*{Acknowledgements}
We are grateful to the referee for the valuable comments and suggestions, which have contributed significantly to improving the quality of the paper. JRL is supported by the Austrian Science Fund (FWF): P-34602, Grant DOI: 10.55776/P34602, and PAT 9039323, Grant-DOI 10.55776/PAT9039323, and the National Natural Science Foundation of China (No. 12471023). QY is funded by the European Union (ERC, UNIVERSE PLUS, 101118787). Views and opinions expressed are however those of the author(s) only and do not necessarily reflect those of the European Union or the European Research Council Executive Agency. Neither the European Union nor the granting authority can be held responsible for them.

\section{Dual-conformal-invariant kinematics in $D{=}4$ and $D{=}3$}
\label{sec:$D{=}3$ kinematics limit}

In this section, we describe the restriction of general $D{=}4$
massless kinematics to a $D{=}3$ subspace. We first recall the
spinor-helicity and momentum-twistor conventions that will be used
below; see, for example, \cite{EH13,Hod}.

\subsection{Spinor-helicity variables, dual coordinates, and momentum twistors}

Let
\[
p_i^\mu=(E_i,p_i^1,p_i^2,p_i^3)\in\CC^4,
\qquad \mu=0,1,2,3,
\]
be the momentum of the $i$th particle. We use the mostly-plus
Minkowski metric
\[
\eta_{\mu\nu}=\operatorname{diag}(-1,1,1,1),
\]
so that
\[
p_i\cdot p_j
=\eta_{\mu\nu}p_i^\mu p_j^\nu
=-E_iE_j+p_i^1p_j^1+p_i^2p_j^2+p_i^3p_j^3.
\]
In particular, for a particle of mass $m_i$,
\[
p_i\cdot p_i=-m_i^2.
\]
For a set $A$ of external particles, the corresponding Mandelstam
invariant is
\[
s_A=\left(\sum_{i\in A}p_i\right)^2.
\]

It is convenient to represent a four-vector by a $2\times2$
bispinor. We define
\[
p_{i,\alpha\dot\alpha}
=
p_{i,\mu}(\sigma^\mu)_{\alpha\dot\alpha},
\qquad
p_{i,\mu}=\eta_{\mu\nu}p_i^\nu,
\]
where $\alpha,\dot\alpha=1,2$ are undotted and dotted spinor indices,
respectively, and
\[
\sigma^0=
\begin{pmatrix}1&0\\0&1\end{pmatrix},
\qquad
\sigma^1=
\begin{pmatrix}0&1\\1&0\end{pmatrix},
\qquad
\sigma^2=
\begin{pmatrix}0&-i\\i&0\end{pmatrix},
\qquad
\sigma^3=
\begin{pmatrix}1&0\\0&-1\end{pmatrix}.
\]
Thus
\begin{equation}
\label{eq:four-dimensional-bispinor}
p_{i,\alpha\dot\alpha}
=
\begin{pmatrix}
-E_i+p_i^3 & p_i^1-ip_i^2\\
p_i^1+ip_i^2 & -E_i-p_i^3
\end{pmatrix},
\end{equation}
and
\[
\det(p_{i,\alpha\dot\alpha})=-p_i\cdot p_i.
\]

We raise and lower spinor indices using the antisymmetric tensors
\[
\epsilon^{12}=\epsilon^{\dot1\dot2}=1,
\qquad
\epsilon_{12}=\epsilon_{\dot1\dot2}=-1,
\]
so that, for example,
\[
\lambda^\alpha=\epsilon^{\alpha\beta}\lambda_\beta,
\qquad
\lambda_\alpha=\epsilon_{\alpha\beta}\lambda^\beta,
\qquad
\epsilon^{\alpha\beta}\epsilon_{\beta\gamma}
=\delta^\alpha{}_\gamma,
\]
and similarly for dotted indices.

For a massless particle, $p_i^2=p_i\cdot p_i=0$, and hence
$p_{i,\alpha\dot\alpha}$ has rank one. It can therefore be factorized
as
\begin{equation}
\label{eq:4d-spinor-factorization}
p_{i,\alpha\dot\alpha}
=
\lambda_{i,\alpha}\widetilde{\lambda}_{i,\dot\alpha}.
\end{equation}
In bra-ket notation we write
\[
p_i=|i\rangle[i|.
\]
The angle and square brackets are
\[
\langle a,b\rangle
=
\epsilon^{\alpha\beta}
\lambda_{a,\alpha}\lambda_{b,\beta},
\qquad
[a,b]
=
\epsilon^{\dot\alpha\dot\beta}
\widetilde{\lambda}_{a,\dot\alpha}
\widetilde{\lambda}_{b,\dot\beta}.
\]
With these conventions,
\begin{equation}
\label{eq:4d-spinor-dot-product}
2p_a\cdot p_b=\langle a,b\rangle[a,b].
\end{equation}
In particular, for two massless momenta,
\[
s_{a,b}=(p_a+p_b)^2=\langle a,b\rangle[a,b].
\]

The factorization \eqref{eq:4d-spinor-factorization} is invariant
under the little-group rescaling
\[
\lambda_{i,\alpha}\longmapsto t_i\lambda_{i,\alpha},
\qquad
\widetilde{\lambda}_{i,\dot\alpha}
\longmapsto t_i^{-1}\widetilde{\lambda}_{i,\dot\alpha},
\qquad
t_i\in\CC^\times.
\]
The massless Weyl equations are
\[
p_{i,\alpha\dot\alpha}\widetilde{\lambda}_i^{\dot\alpha}=0,
\qquad
\lambda_i^\alpha p_{i,\alpha\dot\alpha}=0,
\]
or, in bra-ket notation,
\[
p_i|i]=0,
\qquad
\langle i|p_i=0.
\]

Momentum conservation,
\[
\sum_{i=1}^n p_i^\mu=0,
\]
can be incorporated by introducing dual coordinates $x_i^\mu$ through
\begin{equation}
\label{eq:dual-coordinate-definition}
p_i^\mu=x_{i+1}^\mu-x_i^\mu,
\qquad
x_{n+i}^\mu=x_i^\mu.
\end{equation}
We write
\[
x_{i,j}^\mu=x_i^\mu-x_j^\mu,
\qquad
x_{i,j}^2
=
\eta_{\mu\nu}x_{i,j}^\mu x_{i,j}^\nu.
\]
It follows that
\[
s_{i,i+1,\ldots,j-1}
=
(p_i+p_{i+1}+\cdots+p_{j-1})^2
=
x_{i,j}^2.
\]

As for momenta, the dual coordinates may be represented by
$2\times2$ bispinors. We set
\[
(x_{i,j})^{\dot\alpha\beta}
=
\epsilon^{\dot\alpha\dot\gamma}
\epsilon^{\beta\gamma}
(x_{i,j})_{\gamma\dot\gamma}.
\]
We will use the notation
\[
\langle i|x_{i,k}x_{k,j}|j\rangle
:=
\lambda_i^\alpha
(x_{i,k})_{\alpha\dot\alpha}
(x_{k,j})^{\dot\alpha\beta}
\lambda_{j,\beta}.
\]

Following the momentum-twistor construction \cite{Hod}, we introduce
\[
\mathbf Z_i
=
\left(\lambda_i^\alpha,\mu_i^{\dot\alpha}\right)
\in \mathbb{CP}^3,
\qquad
\mu_i^{\dot\alpha}
=
x_i^{\alpha\dot\alpha}\lambda_{i,\alpha}.
\]
Equivalently,
\[
\mathbf Z_i
=
\left(
\lambda_i^1,\lambda_i^2,
\mu_i^{\dot1},\mu_i^{\dot2}
\right).
\]
The momentum twistors $\mathbf Z_1,\ldots,\mathbf Z_n$ form the
columns of a $4\times n$ matrix. Modulo the $\GL(4)$ action, this
matrix determines a point of $\Gr(4,n)$; quotienting further by the
individual projective rescalings of the momentum twistors gives
\[
\Gr(4,n)/(\CC^\times)^{n-1}.
\]
The corresponding number of independent degrees of freedom is
\[
4(n-4)-(n-1)=3n-15.
\]

We will repeatedly use the standard momentum-twistor identities
\cite{EH13}
\begin{equation}
\label{eq:xij squre is P over two brackets}
x_{i,j}^2
=
\frac{P_{i-1,i,j-1,j}}
{\langle i-1,i\rangle\langle j-1,j\rangle}
\end{equation}
and
\begin{equation}
\label{eq: left i middle xik xkj right j equal to P over one bracket}
\langle i|x_{i,k}x_{k,j}|j\rangle
=
\frac{P_{i,k-1,k,j}}
{\langle k-1,k\rangle}.
\end{equation}
Here
\[
P_{a,b,c,d}
=
\det(\mathbf Z_a,\mathbf Z_b,\mathbf Z_c,\mathbf Z_d)
\]
denotes a Pl\"ucker coordinate. Particle labels are understood
cyclically, and we use the antisymmetry of Pl\"ucker coordinates when
their indices are reordered.

\subsection{$D{=}3$ kinematics}

We now impose the $D=3$ reduction of the four-dimensional kinematics
described above. We regard the resulting kinematic subspace as the
three-dimensional subspace of four-dimensional Minkowski space
obtained by setting the second spatial component of the momentum,
$p_i^{\,2}$, to zero. Thus
\[
p_i^\mu=(E_i,p_i^1,0,p_i^3),
\]
and \eqref{eq:four-dimensional-bispinor} becomes
\[
p_{i,\alpha\dot\alpha}
=
\begin{pmatrix}
-E_i+p_i^3 & p_i^1\\
p_i^1 & -E_i-p_i^3
\end{pmatrix}.
\]
This matrix is symmetric. Under restriction to the three-dimensional
Lorentz subgroup, the dotted and undotted two-component spinor
representations can be identified. We therefore use a single type of
spinor index and write the three-dimensional momentum as a symmetric
bispinor $p_i^{\alpha\beta}$.

For a massless momentum this symmetric matrix has rank one. We choose
the normalization of the spinor so that
\begin{equation}
\label{ptolambda}
p_i^{\alpha\beta}
=
\lambda_i^\alpha\lambda_i^\beta.
\end{equation}
The three-dimensional spinor bracket is
\[
\langle a,b\rangle
=
\epsilon_{\alpha\beta}
\lambda_a^\alpha\lambda_b^\beta.
\]
With the mostly-plus metric used above,
\begin{equation}
\label{eq:D3-dot-product}
2p_a\cdot p_b
=
-\langle a,b\rangle^2.
\end{equation}
Consequently,
\[
x_{a,a+2}^2
=
(p_a+p_{a+1})^2
=
-\langle a,a+1\rangle^2.
\]
Combining this with
\eqref{eq:xij squre is P over two brackets} gives
\begin{equation}
\label{22}
P_{a-1,a,a+1,a+2}
=
-\langle a-1,a\rangle
 \langle a,a+1\rangle^2
 \langle a+1,a+2\rangle.
\end{equation}

We now translate \eqref{ptolambda} into relations among
momentum-twistor Pl\"ucker coordinates. We first record some notation.
By \eqref{eq:dual-coordinate-definition},
\begin{equation}
\label{eq:telescoping-momenta}
p_m+p_{m+1}+\cdots+p_s
=
x_{s+1}-x_m
=
x_{s+1,m}
=
-x_{m,s+1},
\end{equation}
with the same identity understood cyclically. In particular,
\[
p_{a+2}+\cdots+p_{c-1}=x_{c,a+2},
\qquad
p_{c+2}+\cdots+p_{a-1}=x_{a,c+2}.
\]
Notice also that
\[
x_{c+2,c}=p_c+p_{c+1},
\qquad
x_{c,c+1}=-p_c,
\]
and similarly
\[
x_{a+2,a}=p_a+p_{a+1},
\qquad
x_{a,a+1}=-p_a.
\]

All $x_{i,j}$ are symmetric bispinors in the three-dimensional
subspace. For a $2\times2$ matrix $M^{\alpha\beta}$ we write
\[
\langle u|M|v\rangle
:=
\lambda_u^\alpha
\epsilon_{\alpha\gamma}
M^{\gamma\delta}
\epsilon_{\delta\beta}
\lambda_v^\beta,
\]
and similarly
\[
\langle u|MN|v\rangle
:=
\lambda_u^\alpha
\epsilon_{\alpha\gamma}
M^{\gamma\delta}
\epsilon_{\delta\rho}
N^{\rho\sigma}
\epsilon_{\sigma\beta}
\lambda_v^\beta.
\]
If $M$ is symmetric, then
\[
\langle u|M|v\rangle=\langle v|M|u\rangle.
\]
Moreover, \eqref{ptolambda} immediately implies
\begin{equation}
\label{eq:rank-one-sandwich}
\langle u|p_rM|v\rangle
=
\langle u,r\rangle\langle r|M|v\rangle,
\qquad
\langle u|Mp_r|v\rangle
=
\langle u|M|r\rangle\langle r,v\rangle.
\end{equation}
In particular,
\[
p_r|r\rangle=0,
\qquad
\langle r|p_r=0.
\]

For symmetric $2\times2$ matrices $p_1,\ldots,p_k$, define the cyclic
contraction
\begin{equation}
\label{eq:cyclic-contraction-definition}
[[p_1,p_2,\ldots,p_k]]
=
\sum_{\alpha_i,\beta_i=1}^2
p_1^{\alpha_1\beta_1}
p_2^{\alpha_2\beta_2}\cdots
p_k^{\alpha_k\beta_k}
\epsilon_{\alpha_1\beta_k}
\epsilon_{\alpha_2\beta_1}\cdots
\epsilon_{\alpha_k\beta_{k-1}}.
\end{equation}
For general $n$-point kinematics, consider
\[
S=
[[p_a,p_{a+1},(p_{a+2}+\cdots+p_{c-1}),
p_c,p_{c+1},(p_{c+2}+\cdots+p_{a-1})]],
\]
where the particle labels are understood cyclically and
\[
c-a=i+3,
\qquad
i\in\{0,\ldots,n-6\}.
\]
Set
\[
A=x_{c,a+2}=p_{a+2}+\cdots+p_{c-1},
\qquad
B=x_{a,c+2}=p_{c+2}+\cdots+p_{a-1}.
\]
Substituting
$p_j^{\alpha\beta}=\lambda_j^\alpha\lambda_j^\beta$
for $j=a,a+1,c,c+1$ into
\eqref{eq:cyclic-contraction-definition} gives
\begin{equation}
\label{eq:S-factorized}
S
=
\langle a,a+1\rangle
\langle c,c+1\rangle
\langle a+1|A|c\rangle
\langle a|B|c+1\rangle.
\end{equation}
Indeed,
\[
\epsilon_{\alpha_2\beta_1}
\lambda_{a+1}^{\alpha_2}\lambda_a^{\beta_1}
=-\langle a,a+1\rangle,
\qquad
\epsilon_{\alpha_5\beta_4}
\lambda_{c+1}^{\alpha_5}\lambda_c^{\beta_4}
=-\langle c,c+1\rangle,
\]
so these two signs cancel, while the remaining contractions give
the two matrix elements in \eqref{eq:S-factorized}.

We first evaluate $S$ using the pair $p_c,p_{c+1}$. Since
$x_{c+2,c}=p_c+p_{c+1}$ and $\langle c|p_c=0$,
\eqref{eq:rank-one-sandwich} gives
\[
\langle c|x_{c+2,c}B|a\rangle
=
\langle c,c+1\rangle
\langle c+1|B|a\rangle
=
\langle c,c+1\rangle
\langle a|B|c+1\rangle.
\]
Since $A=x_{c,a+2}=-x_{a+2,c}$, it follows from
\eqref{eq:S-factorized} that
\begin{equation}
\label{eq:S-first-intermediate}
S
=
-\langle c|x_{c+2,c}x_{a,c+2}|a\rangle
\langle a,a+1\rangle
\langle a+1|x_{a+2,c}|c\rangle.
\end{equation}
On the other hand, since $x_{c,c+1}=-p_c$,
\[
\langle a+1|x_{a+2,c}x_{c,c+1}|c+1\rangle
=
-\langle c,c+1\rangle
\langle a+1|x_{a+2,c}|c\rangle.
\]
Therefore
\begin{align}
S
&=
\frac{
\langle c|x_{c+2,c}x_{a,c+2}|a\rangle
\langle a,a+1\rangle
\langle a+1|x_{a+2,c}x_{c,c+1}|c+1\rangle
}{
\langle c,c+1\rangle
}
\nonumber\\
&=
\frac{
P_{c,c+1,c+2,a}\langle a,a+1\rangle
P_{a+1,c-1,c,c+1}
}{
\langle c-1,c\rangle
\langle c,c+1\rangle
\langle c+1,c+2\rangle
}.
\label{eq:S-first-evaluation}
\end{align}
For the last equality, we used
\[
x_{c+2,c}=-x_{c,c+2},
\qquad
x_{a,c+2}=-x_{c+2,a},
\]
so that
\[
\langle c|x_{c+2,c}x_{a,c+2}|a\rangle
=
\frac{P_{c,c+1,c+2,a}}
{\langle c+1,c+2\rangle}.
\]
We also used
\[
x_{a+2,c}=x_{a+1,c}+p_{a+1},
\qquad
\langle a+1|p_{a+1}=0,
\]
and hence
\[
\langle a+1|x_{a+2,c}x_{c,c+1}|c+1\rangle
=
\frac{P_{a+1,c-1,c,c+1}}
{\langle c-1,c\rangle},
\]
by
\eqref{eq: left i middle xik xkj right j equal to P over one bracket}.

We next evaluate the same cyclic contraction using
$p_a,p_{a+1}$. Since
$x_{a+2,a}=p_a+p_{a+1}$ and $\langle a|p_a=0$,
\[
\langle a|x_{a+2,a}A|c\rangle
=
\langle a,a+1\rangle
\langle a+1|A|c\rangle.
\]
Moreover, $B=x_{a,c+2}=-x_{c+2,a}$, and the symmetry of the
three-dimensional bispinors gives
\[
\langle a|B|c+1\rangle
=
-\langle c+1|x_{c+2,a}|a\rangle.
\]
Thus \eqref{eq:S-factorized} becomes
\[
S
=
-\langle a|x_{a+2,a}x_{c,a+2}|c\rangle
\langle c,c+1\rangle
\langle c+1|x_{c+2,a}|a\rangle.
\]
Since
\[
x_{c+2,a}=x_{c+1,a}+p_{c+1},
\qquad
\langle c+1|p_{c+1}=0,
\]
we may replace $x_{c+2,a}$ by $x_{c+1,a}$ in the last matrix
element. Furthermore, $x_{a,a+1}=-p_a$ gives
\[
\langle c+1|x_{c+1,a}x_{a,a+1}|a+1\rangle
=
-\langle a,a+1\rangle
\langle c+1|x_{c+1,a}|a\rangle.
\]
Consequently,
\begin{align}
S
&=
\frac{
\langle a|x_{a+2,a}x_{c,a+2}|c\rangle
\langle c,c+1\rangle
\langle c+1|x_{c+1,a}x_{a,a+1}|a+1\rangle
}{
\langle a,a+1\rangle
}
\nonumber\\
&=
\frac{
P_{a,a+1,a+2,c}\langle c,c+1\rangle
P_{c+1,a-1,a,a+1}
}{
\langle a-1,a\rangle
\langle a,a+1\rangle
\langle a+1,a+2\rangle
}.
\label{eq:S-second-evaluation}
\end{align}
Here
\[
x_{a+2,a}=-x_{a,a+2},
\qquad
x_{c,a+2}=-x_{a+2,c},
\]
so that
\[
\langle a|x_{a+2,a}x_{c,a+2}|c\rangle
=
\frac{P_{a,a+1,a+2,c}}
{\langle a+1,a+2\rangle},
\]
while
\[
\langle c+1|x_{c+1,a}x_{a,a+1}|a+1\rangle
=
\frac{P_{c+1,a-1,a,a+1}}
{\langle a-1,a\rangle}.
\]

Equating \eqref{eq:S-first-evaluation} and
\eqref{eq:S-second-evaluation}, and using \eqref{22} both for $a$
and for $c$, gives
\[
P_{a,c,c+1,c+2}
P_{a+1,c-1,c,c+1}
P_{a-1,a,a+1,a+2}
=
P_{a,a+1,a+2,c}
P_{a-1,a,a+1,c+1}
P_{c-1,c,c+1,c+2}.
\]
Here we have used the antisymmetry of the Pl\"ucker coordinates,
\[
P_{c,c+1,c+2,a}
=
-P_{a,c,c+1,c+2},
\qquad
P_{c+1,a-1,a,a+1}
=
-P_{a-1,a,a+1,c+1},
\]
and the two minus signs cancel.

Equivalently, dividing both sides by the appropriate consecutive
Pl\"ucker coordinates, we obtain the projectively invariant relation
\begin{equation}
\label{eq:equations obtained from constrains}
\frac{
P_{a,a+1,a+2,c}
P_{a-1,a,a+1,c+1}
}{
P_{a-1,a,a+1,a+2}
P_{a,a+1,c,c+1}
}
=
\frac{
P_{a+1,c-1,c,c+1}
P_{a,c,c+1,c+2}
}{
P_{c-1,c,c+1,c+2}
P_{a,a+1,c,c+1}
},
\end{equation}
where
\[
c-a=i+3,
\qquad
i\in\{0,\ldots,n-6\}.
\]
Thus restricting the four-dimensional kinematics to the
three-dimensional subspace gives
\eqref{eq:equations obtained from constrains} as a relation among
momentum-twistor Pl\"ucker coordinates.
Naively, there are
$\frac{n(n-5)}{2}$
such relations. However, according to the counting of independent
dual-conformally invariant kinematic variables in $D=3$
\cite{HLY26}, only $n-5$ of them are independent.

In the following sections, we show that these independent
three-dimensional kinematic constraints arise as folding conditions
for suitable seeds of the Grassmannian cluster algebra
$\CC[\Gr(4,n)]$.

\section{Grassmannian cluster algebras} \label{sec:Grassmannian cluster algebras}
In this section, we recall results of Grassmannian cluster algebras \cite{FZ02, Sco, CDFL}. 

\subsection{Cluster algebras}
Cluster algebras were introduced by Fomin and Zelevinsky \cite{FZ02}. We recall the definition. 

For $a \le b \in \ZZ$, we denote $[a,b]=\{a,a+1,\ldots,b\}$. For $a \in \ZZ_{\ge 1}$, we denote $[a] = [1,a]$. 

A quiver $Q=(Q_0, Q_1, s, t)$ is a finite directed graph without loops or $2$-cycles, with vertex set $Q_0$, arrow set $Q_1$, and with maps $s,t: Q_1 \to Q_0$ taking an arrow to its source and target, respectively.

Let $\mathcal{F}$ be an ambient field abstractly isomorphic to a field of rational functions in $m$ independent variables. A seed in $\mathcal{F}$ is a pair $({\bf x}, Q)$, where ${\bf x} = (x_1, \ldots, x_m)$ form a free generating set of $\mathcal{F}$ and $Q$ is a quiver. The set ${\bf x}$ is called the cluster of the seed $({\bf x}, Q)$. The variables $x_1, \ldots, x_n$ are called cluster variables for this seed, and the variables $x_{n+1}, \ldots, x_m$ are called frozen variables. 

For a seed $({\bf x}, Q)$ and $k \in [n]$, the mutated seed $\mu_k({\bf x}, Q)$ is  $({\bf x}', \mu_k(Q))$, where ${\bf x}' = (x_1', \ldots, x_m')$ with $x_j'=x_j$ for $j\ne k$, $x_k' \in \mathcal{F}$ determined by
\begin{align*}
x_k' x_k = \prod_{\alpha \in Q_1, s(\alpha)=k} x_{t(\alpha)} + \prod_{\alpha \in Q_1, t(\alpha)=k} x_{s(\alpha)},
\end{align*} 
and the mutated quiver $\mu_k(Q)$ is a quiver obtained from $Q$ as follows:
\begin{enumerate}
\item[(i)] for each sub-quiver $i \to k \to j$, add a new arrow $i \to j$,

\item[(ii)] reverse the orientation of every arrow with target or source equal to $k$,

\item[(iii)] remove the arrows in a maximal set of pairwise disjoint $2$-cycles. 
\end{enumerate}

The mutation class of a seed $({\bf x}, Q)$ is the set of all seeds obtained from $({\bf x}, Q)$ by a finite sequence of mutations. If $({\bf x}', Q')$ is a seed in the mutation class, then the set ${\bf x}'$ is called a cluster and its elements are called cluster variables. The cluster algebra $\mathcal{A}_{{\bf x}, Q}$ is the subring of $\mathcal{F}$ generated by all cluster variables and frozen variables.

At each mutable vertex $k$, there is a cluster $X$-coordinate $\hat{y}_k = \frac{\prod_{j \to k} x_j}{\prod_{k \to j} x_j}$. 

\subsection{Grassmannian cluster algebras} \label{subsec:grassmannian cluster algebras initial seed}

For $k\le n$, denote by $\Gr(k,n)$ (the affine cone over) the Grassmannian of $k$-dimensional subspaces in $\CC^n$. Elements in $\Gr(k,n)$ can be identified with a $k \times n$ full rank matrix up to row operations. A Pl\"{u}cker coordinate $P_{i_1,\ldots,i_k}$, $1 \le i_1<\ldots <i_k \le n$ on $\Gr(k,n)$ is a regular function sending a matrix $x \in \Gr(k,n)$ to the determinant of the submatrix of $x$ consisting of $1$st, $\ldots$, $k$th rows, and $i_1$th, $\ldots$, $i_k$th columns. 

Denote by $\CC[\Gr(k,n)]$ the coordinate ring of $\Gr(k,n)$. It is generated by the Pl\"{u}cker coordinates $P_{i_1, \ldots, i_{k}}$, $1 \leq i_1 < \cdots < i_{k} \leq n$, subject to the so-called Pl\"ucker relations, see e.g.~\cite{GH14} for more details. 

\begin{figure}
    \centering
    \adjustbox{scale=0.6,center}{
\begin{tikzcd}
	{\fbox{$\begin{matrix} 1 \\ 2 \\ 3 \\4 \end{matrix}$} \ (0,0)}  \\
	\\
	{\begin{matrix} 1 \\ 2 \\ 3 \\5 \end{matrix} \ (1,1) \ (1)} && {\begin{matrix} 1 \\ 2 \\ 4 \\5 \end{matrix} \ (1,2) \ (5)} && {\begin{matrix} 1 \\ 3 \\ 4 \\5 \end{matrix} \ (1,3) \ (9)} && {\fbox{$\begin{matrix} 2 \\ 3 \\ 4 \\5 \end{matrix}$}} \ (1,4) \\
	\\
	{\begin{matrix} 1 \\ 2 \\ 3 \\6 \end{matrix} \ (2,1) \ (2)} && {\begin{matrix} 1 \\ 2 \\ 5 \\6 \end{matrix}  \ (2,2) \ (6)} && {\begin{matrix} 1 \\ 4 \\ 5 \\6 \end{matrix}  \ (2,3)\ (10)} && {\fbox{$\begin{matrix} 3 \\ 4 \\ 5 \\6 \end{matrix}$}}  \ (2,4) \\
	\\
	{\begin{matrix} 1 \\ 2 \\ 3 \\7 \end{matrix}  \ (3,1) \ (3)} && {\begin{matrix} 1 \\ 2 \\ 6 \\7 \end{matrix} \ (3,2)\ (7)} && {\begin{matrix} 1 \\ 5 \\ 6 \\7 \end{matrix} \ (3,3)\ (11)} && {\fbox{$\begin{matrix} 4 \\ 5 \\ 6 \\7 \end{matrix}$}}  \ (3,4) \\
	\\
	{\begin{matrix} 1 \\ 2 \\ 3 \\8 \end{matrix} \ (4,1)\ (4)} && {\begin{matrix} 1 \\ 2 \\ 7 \\8 \end{matrix} \ (4,2) \ (8)} && {\begin{matrix} 1 \\ 6 \\ 7 \\8 \end{matrix}\ (4,3) \ (12) } && {\fbox{$\begin{matrix} 5 \\ 6 \\ 7 \\8 \end{matrix}$}} \ (4,4) \\
	\\
	{\fbox{$\begin{matrix} 1 \\ 2 \\ 3 \\9 \end{matrix}$}} \ (5,1) && {\fbox{$\begin{matrix} 1 \\ 2 \\ 8 \\9 \end{matrix}$}}  \ (5,2) && {\fbox{$\begin{matrix} 1 \\ 7 \\ 8 \\9 \end{matrix}$}}  \ (5,3) && {\fbox{$\begin{matrix} 6 \\ 7 \\ 8 \\9 \end{matrix}$}}  \ (5,4)
	\arrow[from=3-1, to=1-1]
	\arrow[from=5-1, to=3-1]
	\arrow[from=3-1, to=5-3]
	\arrow[from=3-3, to=3-1]
	\arrow[from=3-5, to=3-3]
	\arrow[from=3-3, to=5-5]
	\arrow[from=5-3, to=3-3]
	\arrow[from=5-5, to=3-5]
	\arrow[from=3-5, to=5-7]
	\arrow[from=3-7, to=3-5]
	\arrow[from=5-7, to=5-5]
	\arrow[from=5-5, to=5-3]
	\arrow[from=5-3, to=5-1]
	\arrow[from=5-1, to=7-3]
	\arrow[from=7-1, to=5-1]
	\arrow[from=7-3, to=7-1]
	\arrow[from=9-1, to=7-1]
	\arrow[from=9-3, to=7-3]
	\arrow[from=7-3, to=5-3]
	\arrow[from=5-3, to=7-5]
	\arrow[from=7-5, to=7-3]
	\arrow[from=7-5, to=5-5]
	\arrow[from=5-5, to=7-7]
	\arrow[from=7-7, to=7-5]
	\arrow[from=7-5, to=9-7]
	\arrow[from=9-5, to=7-5]
	\arrow[from=7-3, to=9-5]
	\arrow[from=7-1, to=9-3]
	\arrow[from=11-1, to=9-1]
	\arrow[from=9-3, to=9-1]
	\arrow[from=9-5, to=9-3]
	\arrow[from=9-7, to=9-5]
	\arrow[from=9-5, to=11-7]
	\arrow[from=11-5, to=9-5]
	\arrow[from=11-3, to=9-3]
	\arrow[from=9-3, to=11-5]
	\arrow[from=9-1, to=11-3]
\end{tikzcd}
}
\caption{An initial seed for $\Gr(4,9)$. We label the mutable vertices as $(1), (2), \ldots, (12)$.}
\label{fig:initial seed Gr49}
\end{figure}
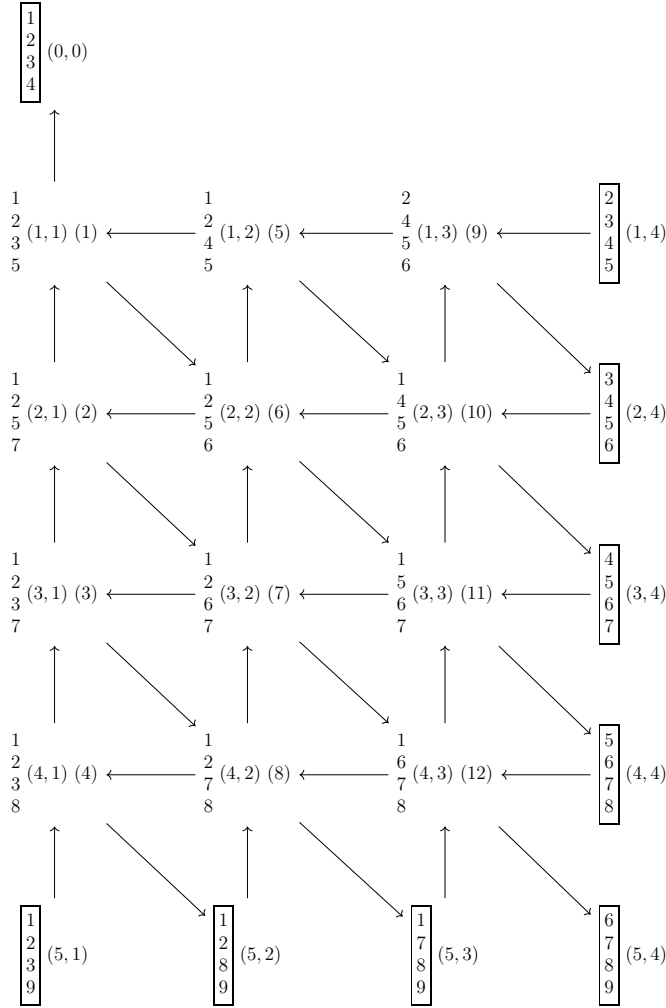

Scott \cite{Sco} proved that the coordinate ring $\CC[\Gr(k,n)]$ has a cluster algebra structure. The cluster algebra $\CC[\Gr(k,n)]$ has an initial seed $({\bf x}, Q)$ with the initial quiver $Q$ with vertices 
\[
\{(0,0)\} \cup \{(a,b) : a\in [n-k], \ b\in [k]
\}
\]
and arrows 
\[
\begin{array}{rl}
(0,0)\to (1,1), \\
(a-1,b)\to (a,b), &  a\in [2,n-k],\ b\in [k], \\
(a,b-1)\to (a,b), &  a\in [n-k],\ b\in [2,k], \\
(a+1,b+1)\to (a,b), &  a\in [n-k-1],\ b\in [k-1]. 
\end{array}
\] 
The quiver in Figure \ref{fig:initial seed Gr49} is the initial quiver of $\CC[\Gr(4,9)]$. 

The cluster variables (and frozen variables) in this initial seed are certain Pl\"{u}cker coordinates. The frozen variable at $(0,0)$ is $P_{1,\ldots,k}$. The cluster variables (including frozen variables) in the column with $b=1$ are $P_{1,2,\ldots, k-1, k+1}$, $\ldots$, $P_{1,2,\ldots, k-1, n}$. The cluster variables (including frozen variables) in column with $b=2$ are $P_{1,2,\ldots, k-2, k, k+1}$, $\ldots$, $P_{1,2,\ldots, k-2, n-1, n}$. The column with $b=k$ consists of frozen variables $P_{2,\ldots, k+1}$, $\ldots$, $P_{n-k+1, \ldots, n}$. Figure \ref{fig:initial seed Gr49} is the case of $\Gr(4,9)$.


\subsection{Mutations in Grassmannian cluster algebras in terms of tableaux} \label{subsec:Mutation of Grassmannian cluster algebras in terms of tableaux}
A Young diagram (also called Ferrers diagram) is a graphical representation of an integer partition $\lambda = (\lambda_1 \ge \lambda_2 \ge \cdots \ge \lambda_l\geq 0)$. The Young diagram of the partition $\lambda$ has $\lambda_i$ boxes in the $i$th row. The boxes are adjusted to the north-west in the $4$th quadrant of a $2$-dimensional Cartesian coordinate system. A Young tableau is a labelling of the boxes of a Young diagram with positive natural numbers. A semistandard Young tableau is a Young tableau where the entries are weakly increasing in each row and strictly increasing in each column. For Grassmannian cluster algebras, we only need to use semistandard Young tableaux of rectangular shapes. For $k \le n \in \ZZ_{\ge 1}$, we denote by ${\rm SSYT}(k, [n])$ the set of rectangular semistandard Young tableaux with $k$ rows and with entries in $[n]$ (with arbitrarily many columns). 

For $S, T \in {\rm SSYT}(k, [n])$, let $S \cup T$ be the row-increasing tableau whose $i$th row is the union of the $i$th rows of $S$ and $T$ (as multisets), for any $i$, \cite{CDFL}. 
By Lemma 3.6 in \cite{CDFL}, $S \cup T$ is in ${\rm SSYT}(k, [n])$. We call $S$ a factor of $T$, and write $S \subset T$, if the $i$th row of $S$ is contained in that of $T$ (as multisets), for every $i \in [k]$. In this case, we define $\frac{T}{S}=S^{-1}T=T S^{-1}$ to be the row-increasing tableau whose $i$th row is obtained by removing that of $S$ from that of $T$ (as multisets), for every $i \in [k]$. 

Every element in the dual canonical basis (in particular, every cluster variable) of $\CC[\Gr(k,n)]$ corresponds to a tableau in $\SSYT(k,[n])$, see \cite[Section 3]{CDFL}. Denote by $\ch(T)$ the dual canonical basis element corresponding to $T \in \SSYT(k,[n])$. 

There is a partial order called dominance order in the set of semi-standard Young tableaux \cite[Section 5.5]{Bri05}. 
Let $\lambda = (\lambda_1,\dots,\lambda_l)$, $\mu = (\mu_1,\dots,\mu_l)$, with $\lambda_1 \geq \cdots \geq \lambda_l \geq 0$, $\mu_1 \ge \cdots \ge \mu_l \ge 0$, be partitions. 
Then 
\[
\lambda \le \mu\quad  \text{  in the dominance order if } \quad  \sum_{j \leq i}\lambda_j \le \sum_{j \leq i}\mu_j \text{ for all } 1\le i\le l.
\] 
For $T\in {\rm SSYT}(k,[n])$ and $i \in [n]$, denote by $T[i]$ the sub-tableau obtained from $T$ by restriction to the entries in $[i]$. 
For a tableau $T$, let ${\rm sh}(T)$ denote the shape of $T$. 
If $T,T' \in {\rm SSYT}(k,[n])$ are of the same shape, then $T \le T'$ in the dominance order if for every $i \in [n]$, ${\rm sh}(T[i]) \le {\rm sh}({T'}[i])$ in the dominance order on partitions. 

Mutations of cluster variables in the cluster algebra $\CC[\Gr(k,n)]$ can be described in terms of tableaux \cite[Section 4]{CDFL}. 
Starting from an initial seed of $\CC[\Gr(k,n)]$, each time we perform a mutation at a cluster variable $\ch(T_r)$, we obtain a new cluster variable $\ch(T'_r)$ determined by
\begin{align}\label{eq:char mut}
\ch(T'_r)\ch(T_r) = \prod_{i \to r} \ch(T_i) + \prod_{r \to i} \ch(T_i),
\end{align}
where $\ch(T_i)$ is the cluster variable at the vertex $i$. The two tableaux $\cup_{i \to r} T_i$, $\cup_{r \to i} T_i$ are always comparable under the dominance order and $T'_r$ is determined by
\begin{align}\label{eq:mutation tableaux}
T'_r = T^{-1}_r \max\{\cup_{i \to r} T_i, \cup_{r \to i} T_i \}. 
\end{align}

\section{Folding of Grassmannian cluster algebras}
\label{sec:folding of Grkn}

In this section, we show that the kinematic constraints restricting
general $D{=}4$ kinematics to the $D{=}3$ subspace obtained in
Section~\ref{sec:$D{=}3$ kinematics limit} can be realized as folding
conditions on Grassmannian cluster algebras. We first construct, for
$\CC[\Gr(2r,n)]$, a seed whose mutable quiver admits a natural
reflection symmetry. We then specialize to $\Gr(4,n)$ and prove that
the corresponding folding equations are precisely the kinematic
constraints obtained in
Section~\ref{sec:$D{=}3$ kinematics limit}.

For the purposes of this paper, we use the following notion of folding.

\begin{definition}
\label{def:folding}
Let $(\mathbf{x},Q)$ be a seed and let $Q_{\mathrm{mut}}$ be its
mutable subquiver. Let $\sigma$ be an involution of the mutable
vertices. We say that $Q_{\mathrm{mut}}$ is \emph{foldable with
respect to $\sigma$} if $\sigma$ is an automorphism of
$Q_{\mathrm{mut}}$ and no two distinct vertices in the same
$\sigma$-orbit are joined by an arrow. The corresponding
\emph{folding conditions} are
\[
X_v=X_{\sigma(v)}
\]
for all mutable vertices $v$; for a fixed point of $\sigma$ this
condition is vacuous.
\end{definition}

We emphasize that only these folding conditions on cluster
$X$-coordinates will be used below; we do not need to endow the
quotient quiver with a separate cluster algebra structure.

\subsection{A foldable seed for $\Gr(2r,n)$}
\label{subsec:folable seed for Gr2rn}

Let $k=2r$, with $r\geq2$, and let $n\geq k+2$. Set
\[
\ell=n-k-1.
\]
Let $(\mathbf{x},Q)$ be the standard initial seed of
$\CC[\Gr(k,n)]$ described in
Section~\ref{subsec:grassmannian cluster algebras initial seed}.
We index its mutable vertices by
\[
v_{a,b}=(a,b),
\qquad
1\leq a\leq\ell,\qquad
1\leq b\leq k-1,
\]
where $a$ increases from top to bottom and $b$ from left to right.

For $1\leq b\leq k-1$ and $j\geq1$, let $C_{b,j}$ denote the
sequence of mutations
\[
v_{1,b},v_{2,b},\ldots,v_{j,b}.
\]
We regard $C_{b,j}$ as empty when $j\leq0$. 
The mutation sequence is organized into successive rounds. 
For $s\geq1$ and $1\leq b\leq k-1$, define
\begin{equation}
\label{eq:hsb}
h_s(b)=\min\{\ell,\ell+b-2s-1\}.
\end{equation}
The $s$-th round of mutations is
\begin{equation}
\label{eq:round-s}
C_{k-1,h_s(k-1)},
C_{k-2,h_s(k-2)},\ldots,
C_{1,h_s(1)},
\end{equation}
where a block $C_{b,h_s(b)}$ is omitted whenever $h_s(b)\leq0$.
Equivalently, the vertices mutated in the $s$-th round are precisely
\begin{equation}
\label{eq:vertices-round-s}
v_{a,b}
\qquad\text{such that}\qquad
b-a\geq2s+1-\ell,
\end{equation}
and they are mutated in decreasing order of $b$ and, for fixed $b$,
in increasing order of $a$.

For example, the first two rounds are
\[
C_{k-1,\ell},C_{k-2,\ell},\ldots,C_{3,\ell},
C_{2,\ell-1},C_{1,\ell-2},
\]
and
\[
C_{k-1,\ell},\ldots,C_{5,\ell},
C_{4,\ell-1},C_{3,\ell-2},
C_{2,\ell-3},C_{1,\ell-4},
\]
respectively, with blocks having non-positive second index omitted. The process terminates when all the blocks in
\eqref{eq:round-s} are empty. Since the largest possible value of
$b-a$ for a mutable vertex is $k-2$, the $s$-th round is nonempty
if and only if
\[
2s+1-\ell\leq k-2.
\]
Hence the last nonempty round is
\begin{equation}
\label{eq:last-round}
S=
\left\lfloor\frac{\ell+k-3}{2}\right\rfloor.
\end{equation}
Thus the full mutation sequence consists of the rounds
$s=1,\ldots,S$. We denote by $(\mathbf{x}',Q')$ the seed obtained
after completing these $S$ rounds.

\begin{theorem}
\label{thm:foldable seed Gr2rn}
Let $k=2r$, $r\geq2$, and $n\geq k+2$. The seed
$(\mathbf{x}',Q')$ obtained by the mutation sequence above has the
following properties.

\begin{enumerate}
\item
The mutable vertices form an $\ell\times(k-1)$ rectangular array,
and there are no arrows between mutable vertices except horizontal
and vertical nearest-neighbor arrows. Every elementary square with
vertices
\[
v_{a,b},\quad v_{a+1,b},\quad
v_{a,b+1},\quad v_{a+1,b+1}
\]
is an oriented $4$-cycle. More precisely, if
$(-1)^{a+b+\ell}=1$, then its orientation is
\begin{equation}
\label{eq:final-square-orientation}
v_{a,b}\longrightarrow v_{a+1,b}
\longrightarrow v_{a+1,b+1}
\longrightarrow v_{a,b+1}
\longrightarrow v_{a,b},
\end{equation}
and if $(-1)^{a+b+\ell}=-1$, all four arrows are reversed.

\item
Let $\rho$ be the twisted cyclic shift of $\Gr(k,n)$, so that
$\rho(P_I)=P_{\rho(I)}$, where $\rho(I)$ is obtained by adding $1$
to every index modulo $n$. Then the final cluster variable at
$v_{a,b}$ is
\begin{equation}
\label{eq:final-cluster-variable}
x'_{a,b}
=
P_{\rho^{N_{a,b}}(I_{a,b})},
\end{equation}
where
\begin{equation}
\label{eq:Iab-and-Nab}
I_{a,b}
=
[1,k-b]\cup[k-b+a+1,k+a],
\qquad
N_{a,b}
=
\left\lfloor
\frac{\ell+b-a-1}{2}
\right\rfloor.
\end{equation}
In particular, every mutable cluster variable in
$(\mathbf{x}',Q')$ is a Pl\"ucker coordinate.

\item
The involution
\begin{equation}
\label{eq:folding-involution}
\sigma(a,b)=(a,k-b)
\end{equation}
is an automorphism of the mutable quiver $Q'$, and no two distinct
vertices in the same $\sigma$-orbit are joined by an arrow.
Consequently, $Q'_{\mathrm{mut}}$ is foldable with respect to
$\sigma$ in the sense of Definition~\ref{def:folding}.
\end{enumerate}
\end{theorem}

\begin{proof}
We write
\[
v_{a,b}=(a,b),
\qquad
1\leq a\leq\ell,\qquad
1\leq b\leq k-1,
\]
for the mutable vertices.

We first determine the mutable part of the quiver. For $s\geq0$,
let $Q^{(s)}$ denote the mutable quiver obtained after the first
$s$ rounds, with $Q^{(0)}$ the mutable part of the standard initial
quiver, and set
\begin{equation}
\label{eq:tau-s}
\tau_s=2s+2-\ell.
\end{equation}
For $1\leq a\leq\ell-1$,
$1\leq b\leq k-2$,
let $\square_{a,b}$ denote the elementary square with vertices
\[
v_{a,b},\qquad
v_{a+1,b},\qquad
v_{a,b+1},\qquad
v_{a+1,b+1}.
\]
When $\ell=1$, there are no elementary squares, and statements below
concerning them are understood vacuously.

We claim inductively that after the first $s$ rounds the only mutable
arrows are horizontal and vertical nearest-neighbor arrows together
with the following diagonals. More precisely, $\square_{a,b}$
contains a diagonal if and only if
\begin{equation}
\label{eq:diagonal-condition}
b-a\geq\tau_s.
\end{equation}
When \eqref{eq:diagonal-condition} holds, the arrows in
$\square_{a,b}$ agree with those of the corresponding square in the
standard initial quiver:
\[
v_{a,b}\longrightarrow v_{a+1,b},
\qquad
v_{a,b}\longrightarrow v_{a,b+1},
\]
\[
v_{a+1,b}\longrightarrow v_{a+1,b+1},
\qquad
v_{a,b+1}\longrightarrow v_{a+1,b+1},
\]
together with the diagonal
\[
v_{a+1,b+1}\longrightarrow v_{a,b}.
\]
If $b-a<\tau_s$, the diagonal is absent and the four boundary arrows
form an oriented $4$-cycle. More precisely, if
$(-1)^{a+b+\ell}=1$, then
\begin{equation}
\label{eq:final-square-orientation}
v_{a,b}\longrightarrow v_{a+1,b}
\longrightarrow v_{a+1,b+1}
\longrightarrow v_{a,b+1}
\longrightarrow v_{a,b},
\end{equation}
while for $(-1)^{a+b+\ell}=-1$ all four arrows are reversed.

For $s=0$, we have $\tau_0=2-\ell$. Every elementary square satisfies
\[
b-a\geq1-(\ell-1)=2-\ell=\tau_0.
\]
Thus every elementary square contains its initial diagonal, and its
arrows agree with those of the standard initial quiver. Hence the
claim holds for $s=0$.

Assume now that the claim holds after the first $s-1$ rounds. Since
\[
\tau_{s-1}=\tau_s-2,
\]
the $s$-th round mutates, by
\eqref{eq:vertices-round-s}, precisely the vertices
\begin{equation}
\label{eq:mutated-region-tau}
v_{a,b}
\qquad\text{with}\qquad
b-a\geq\tau_s-1.
\end{equation}

We examine one mutation in this round. Suppose that
$v=v_{a,b}$ is about to be mutated and, whenever the corresponding
mutable vertices exist, put
\[
U=v_{a-1,b},\qquad
D=v_{a+1,b},\qquad
L=v_{a,b-1},\qquad
R=v_{a,b+1}.
\]
The mutations in a round are performed from right to left in the
columns and from top to bottom within each column. An induction in
this mutation order, using the induction hypothesis on $Q^{(s-1)}$,
shows that immediately before the mutation at $v$ the arrows incident
with $v$ are
\begin{equation}
\label{eq:local-cross}
L\longrightarrow v\longleftarrow R,
\qquad
U\longleftarrow v\longrightarrow D,
\end{equation}
with nonexistent boundary vertices omitted. Moreover, whenever the
vertices involved exist, there are arrows
\begin{equation}
\label{eq:two-local-diagonals}
D\longrightarrow L,
\qquad
D\longrightarrow R.
\end{equation}
Here $D\to R$ is the temporary diagonal created by the mutation
previously performed in the column immediately to the right.

For completeness, this local statement is itself proved by induction
along the order within the $s$-th round. It holds at the first
mutation by the description of $Q^{(s-1)}$. After mutating a vertex,
the calculation below gives exactly the local configuration required
at the next vertex, either one step below in the same column or,
when that column is finished, at the top of the next nonempty column
to the left. Thus \eqref{eq:local-cross} and
\eqref{eq:two-local-diagonals} hold throughout the round.

We now mutate at $v$. The four arrows in
\eqref{eq:local-cross} are reversed. The four length-two paths through
$v$ create
\[
L\longrightarrow U,\qquad
L\longrightarrow D,\qquad
R\longrightarrow U,\qquad
R\longrightarrow D.
\]
The arrows $L\to D$ and $R\to D$ cancel, respectively, with the two
arrows in \eqref{eq:two-local-diagonals}. Hence the only new
diagonals remaining immediately after the mutation are
\[
L\longrightarrow U,
\qquad
R\longrightarrow U.
\]

The first of these is temporary. Indeed, if $L$ and $U$ exist, then
\[
L=v_{a,b-1},
\qquad
U=v_{a-1,b},
\]
and the arrow $L\to U$ is cancelled later in the same round when
$v_{a-1,b-1}$ is mutated. This vertex belongs to the $s$-th round
because
\[
(b-1)-(a-1)=b-a\geq\tau_s-1.
\]
The second arrow, $R\longrightarrow U$,
is the diagonal of the elementary square whose upper-left vertex is
$v_{a-1,b}$. For this square,
\[
b-(a-1)=b-a+1\geq\tau_s,
\]
so this is precisely one of the diagonals that remains after the
$s$-th round.

It remains to determine which old diagonals survive the entire round.
The arrow $D\to L$, removed when $v_{a,b}$ is mutated, is the
diagonal of the square whose upper-left vertex is $v_{a,b-1}$.
That diagonal is recreated later in the same round exactly when
$v_{a+1,b-1}$ is also mutated. By
\eqref{eq:mutated-region-tau}, this happens exactly when
\[
(b-1)-(a+1)\geq\tau_s-1,
\]
or equivalently $(b-1)-a\geq\tau_s$.
Consequently, after completion of the $s$-th round, an elementary
square $\square_{a,b}$ contains a diagonal exactly when
\[
b-a\geq\tau_s.
\]

The same local calculation determines the orientation of the squares
whose diagonals disappear in this round. Since
$\tau_{s-1}=\tau_s-2$, the two new layers are
\[
b-a=\tau_s-2
\qquad\text{and}\qquad
b-a=\tau_s-1.
\]
For the first layer,
\[
a+b+\ell\equiv(\tau_s-2)+\ell=2s\pmod2,
\]
and the resulting orientation is
\eqref{eq:final-square-orientation}; for the second layer,
\[
a+b+\ell\equiv(\tau_s-1)+\ell=2s+1\pmod2,
\]
and all four arrows are reversed. Squares with
$b-a<\tau_s-2$ contain no vertex mutated in the $s$-th round and
therefore remain unchanged. Thus the checkerboard orientation in the
claim is preserved.

Finally, every new arrow created by a length-two path through a
mutated vertex is one of the four arrows listed above. It is therefore
either a horizontal or vertical nearest-neighbor arrow, a diagonal
of an elementary square, or a temporary diagonal cancelled later in
the same round. Hence no other mutable arrows are created. This
completes the induction.

By \eqref{eq:last-round}, after the last round we have
\[
\tau_S=2S+2-\ell>k-3.
\]
On the other hand, every elementary square satisfies
\[
b-a\leq(k-2)-1=k-3.
\]
Hence $b-a<\tau_S$ for every elementary square. By the induction
claim, no elementary square contains a diagonal after the last round.
Therefore the only mutable arrows of $Q'$ are horizontal and vertical
nearest-neighbor arrows, and every elementary square is an oriented
$4$-cycle with the checkerboard orientation
\eqref{eq:final-square-orientation}. This proves part~(1).

We next determine the cluster variables. For
$0\leq a\leq\ell+1$,
$0\leq b\leq k$,
define the $k$-element subset
\begin{equation}
\label{eq:Iab}
I_{a,b}
=
[1,k-b]\cup[k-b+a+1,k+a],
\end{equation}
where an interval is empty if its lower endpoint exceeds its upper
endpoint. For
$1\leq a\leq\ell$ and $1\leq b\leq k-1$, the cluster variable at
$v_{a,b}$ in the standard initial seed is
\begin{equation}
\label{eq:initial-Iab}
x_{a,b}=P_{I_{a,b}}.
\end{equation}
The cases $a=\ell+1$ and $b=k$ give cyclically consecutive
$k$-subsets and hence the frozen Pl\"ucker coordinates on the bottom
and right boundaries. We also have
$I_{0,b}=I_{a,0}=[1,k]$.
The latter are only formal boundary index sets; they will be used to
write the exchange relations uniformly.
Let $\rho$ denote the twisted cyclic shift, normalized so that
\[
\rho(P_I)=P_{\rho(I)},
\]
where $\rho(I)$ is obtained by adding $1$ to every index modulo $n$.
In particular, the cyclic shifts of all the boundary index sets just
described are again cyclically consecutive $k$-subsets and therefore
correspond to frozen Pl\"ucker coordinates.

We record the Pl\"ucker relation governing every mutation. Put $p=k-b$
and
\[
J=[2,p]\cup[p+a+2,k+a].
\]
Then, with empty intervals omitted,
\[
\begin{aligned}
I_{a,b}
&=J\cup\{1,p+a+1\},\\
\rho(I_{a,b})
&=J\cup\{p+1,k+a+1\},\\
I_{a,b-1}
&=J\cup\{1,p+1\},\\
\rho(I_{a,b+1})
&=J\cup\{p+a+1,k+a+1\},\\
\rho(I_{a-1,b})
&=J\cup\{p+1,p+a+1\},\\
I_{a+1,b}
&=J\cup\{1,k+a+1\}.
\end{aligned}
\]
Since
\[
1<p+1<p+a+1<k+a+1,
\]
the standard three-term Pl\"ucker relation gives
\begin{equation}
\label{eq:fundamental-local-plucker}
P_{I_{a,b}}P_{\rho(I_{a,b})}
=
P_{I_{a,b-1}}P_{\rho(I_{a,b+1})}
+
P_{\rho(I_{a-1,b})}P_{I_{a+1,b}}.
\end{equation}
Applying $\rho^q$ gives, for every $q\geq0$,
\begin{equation}
\label{eq:shifted-local-plucker}
P_{\rho^q(I_{a,b})}P_{\rho^{q+1}(I_{a,b})}
=
P_{\rho^q(I_{a,b-1})}
P_{\rho^{q+1}(I_{a,b+1})}
+
P_{\rho^{q+1}(I_{a-1,b})}
P_{\rho^q(I_{a+1,b})}.
\end{equation}

We now prove inductively along the complete mutation sequence that,
whenever $v_{a,b}$ is mutated in round $s$, its cluster variable
changes as
\begin{equation}
\label{eq:shift-at-each-mutation}
P_{\rho^{s-1}(I_{a,b})}
\longmapsto
P_{\rho^s(I_{a,b})}.
\end{equation}
Indeed, by \eqref{eq:vertices-round-s}, the regions mutated in
successive rounds are nested. Thus, if $v_{a,b}$ occurs in round
$s$, then it occurred once in every preceding round and has already
been mutated exactly $s-1$ times. By induction, its current cluster
variable is therefore $P_{\rho^{s-1}(I_{a,b})}$.
Immediately before the mutation at $v_{a,b}$, the vertex to its right
and the vertex above it, when mutable, have already been treated in
the current round, whereas the vertex to its left and the vertex
below it have not. The local quiver calculation used above therefore
shows that the two exchange monomials are
\begin{equation}
\label{eq:exchange-monomials}
P_{\rho^{s-1}(I_{a,b-1})}
P_{\rho^s(I_{a,b+1})}
\qquad\text{and}\qquad
P_{\rho^s(I_{a-1,b})}
P_{\rho^{s-1}(I_{a+1,b})}.
\end{equation}
At a boundary, the symbols involving
$a=0$, $b=0$, $a=\ell+1$, or $b=k$ in
\eqref{eq:exchange-monomials} are interpreted as the corresponding
cyclically consecutive frozen Pl\"ucker coordinates. Inspecting the
boundary arrows of the standard ice quiver and applying the same
local mutation calculation shows that these are exactly the frozen
factors occurring in the exchange relation. Thus
\eqref{eq:exchange-monomials} is valid uniformly at interior and
boundary mutable vertices.
The exchange relation at $v_{a,b}$ is consequently
\[
\begin{aligned}
P_{\rho^{s-1}(I_{a,b})}\,x_{a,b}^{\mathrm{new}}
=
P_{\rho^{s-1}(I_{a,b-1})}
P_{\rho^s(I_{a,b+1})}+
P_{\rho^s(I_{a-1,b})}
P_{\rho^{s-1}(I_{a+1,b})}.
\end{aligned}
\]
By \eqref{eq:shifted-local-plucker}, with $q=s-1$, the right-hand
side equals
\[
P_{\rho^{s-1}(I_{a,b})}
P_{\rho^s(I_{a,b})}.
\]
Cancelling the nonzero factor
$P_{\rho^{s-1}(I_{a,b})}$ in the ambient field gives
\[
x_{a,b}^{\mathrm{new}}
=
P_{\rho^s(I_{a,b})},
\]
proving \eqref{eq:shift-at-each-mutation}.
The number of rounds in which $v_{a,b}$ occurs is
\[
\begin{aligned}
N_{a,b}
&=
\#\{s\geq1:\ b-a\geq2s+1-\ell\}=
\left\lfloor
\frac{\ell+b-a-1}{2}
\right\rfloor.
\end{aligned}
\]
Hence the final cluster variable at $v_{a,b}$ is
\[
x'_{a,b}
=
P_{\rho^{N_{a,b}}(I_{a,b})},
\]
which proves part~(2).

It remains to prove part~(3). Suppose first that $\ell\geq2$.
Reflection in the vertical axis sends the elementary square
$\square_{a,b}$ to
\[
\square_{a,k-b-1}.
\]
Since $k$ is even,
\[
(a+k-b-1+\ell)-(a+b+\ell)
=
k-1-2b
\]
is odd. Thus the checkerboard parity of the reflected square is
opposite to that of the original square. Geometric reflection also
reverses the orientation of a $4$-cycle. These two reversals cancel,
so the map
\[
\sigma(a,b)=(a,k-b)
\]
preserves every mutable arrow. Hence $\sigma$ is an automorphism of
$Q'_{\mathrm{mut}}$.

When $\ell=1$, the mutable quiver is a path. In this case
\eqref{eq:vertices-round-s} says that round $s$ mutates the vertices
\[
v_{1,k-1},v_{1,k-2},\ldots,v_{1,2s+1}.
\]
A direct induction on the rounds gives the final alternating
orientation
\[
v_{1,1}\longrightarrow v_{1,2}
\longleftarrow v_{1,3}
\longrightarrow v_{1,4}
\longleftarrow\cdots.
\]
This path is also invariant under
$b\mapsto k-b$, because $k$ is even. Thus $\sigma$ is a quiver
automorphism for every $\ell\geq1$.
Finally, if $\sigma(a,b)\neq(a,b)$, then
\[
|b-(k-b)|=|k-2b|\geq2,
\]
since $k$ is even. By part~(1), mutable arrows join only vertices in
the same or adjacent columns. Therefore no two distinct vertices in
the same $\sigma$-orbit are joined by an arrow. Hence
$Q'_{\mathrm{mut}}$ is foldable with respect to $\sigma$ in the
sense of Definition~\ref{def:folding}. This proves part~(3), and
completes the proof.
\end{proof}

\subsection{Folding of $\CC[\Gr(4,n)]$}

We now specialize Theorem~\ref{thm:foldable seed Gr2rn} to $k=4$.
Then $\ell=n-5$, and the mutation sequence is
\[
C_{3,\ell},C_{2,\ell-1},C_{1,\ell-2},\quad
C_{3,\ell-2},C_{2,\ell-3},C_{1,\ell-4},\quad
C_{3,\ell-4},C_{2,\ell-5},C_{1,\ell-6},\quad\ldots,
\]
where terms with non-positive second index are omitted. The
involution \eqref{eq:folding-involution} becomes
\[
\sigma(r,1)=(r,3),
\qquad
\sigma(r,2)=(r,2).
\]
Hence the nontrivial folding conditions are
\begin{equation}
\label{eq:X-folding-Gr4n}
X_{r,1}=X_{r,3},
\qquad
1\leq r\leq n-5.
\end{equation}

\begin{corollary}
\label{cor:folding Gr4n}
For the seed $(\mathbf{x}',Q')$ constructed above, the folding
conditions \eqref{eq:X-folding-Gr4n} are precisely the $n-5$
three-dimensional kinematic constraints obtained in
Section~\ref{sec:$D{=}3$ kinematics limit}.
\end{corollary}

\begin{proof}
For $0\leq i\leq n-6$, set
\begin{equation}
\label{eq:ai-ci}
a_i=
\left\lfloor\frac{n-i-3}{2}\right\rfloor,
\qquad
c_i=a_i+i+3.
\end{equation}
Thus $c_i-a_i=i+3$. By
\eqref{eq:final-cluster-variable}, for $1\leq r\leq n-5$,
\begin{equation}
\label{eq:Gr4-final-left-right}
x'_{r,1}
=
P_{a_r,a_r+1,a_r+2,c_r},
\qquad
x'_{r,3}
=
P_{a_r+1,c_r-1,c_r,c_r+1}.
\end{equation}
Indeed,
\[
I_{r,1}=\{1,2,3,r+4\},
\qquad
N_{r,1}=a_r-1,
\]
and
\[
I_{r,3}=\{1,r+2,r+3,r+4\},
\qquad
N_{r,3}=a_r.
\]

Fix $i\in\{0,\ldots,n-6\}$ and put $r=i+1$. Since
\[
a_{i+2}=a_i-1,
\qquad
c_{i+2}=c_i+1,
\]
the oriented-square description in
Theorem~\ref{thm:foldable seed Gr2rn}, together with the frozen
boundary variables, gives
\begin{align}
X_{r,1}
&=
\left(
\frac{
P_{a_i,a_i+1,a_i+2,c_i}
P_{a_i-1,a_i,a_i+1,c_i+1}
}{
P_{a_i-1,a_i,a_i+1,a_i+2}\,x'_{r,2}
}
\right)^{\varepsilon_r},
\label{eq:X-left-Gr4}\\
X_{r,3}
&=
\left(
\frac{
P_{a_i+1,c_i-1,c_i,c_i+1}
P_{a_i,c_i,c_i+1,c_i+2}
}{
P_{c_i-1,c_i,c_i+1,c_i+2}\,x'_{r,2}
}
\right)^{\varepsilon_r},
\label{eq:X-right-Gr4}
\end{align}
where $\varepsilon_r\in\{\pm1\}$ records the checkerboard
orientation. The same formulas apply for $i=0$ and $i=n-6$; the
corresponding cyclic consecutive Pl\"ucker coordinates are frozen
variables.

Since the exponent $\varepsilon_r$ and the factor $x'_{r,2}$ are
the same on both sides, the condition $X_{r,1}=X_{r,3}$ is
equivalent to
\begin{equation}
\label{eq:folding Gr4n}
\frac{
P_{a_i,a_i+1,a_i+2,c_i}
P_{a_i-1,a_i,a_i+1,c_i+1}
}{
P_{a_i-1,a_i,a_i+1,a_i+2}
}
=
\frac{
P_{a_i+1,c_i-1,c_i,c_i+1}
P_{a_i,c_i,c_i+1,c_i+2}
}{
P_{c_i-1,c_i,c_i+1,c_i+2}
}.
\end{equation}
Multiplying the denominators on both sides by the common factor
$P_{a_i,a_i+1,c_i,c_i+1}$, this is exactly
\[
\frac{
P_{a,a+1,a+2,c}
P_{a-1,a,a+1,c+1}
}{
P_{a-1,a,a+1,a+2}
P_{a,a+1,c,c+1}
}
=
\frac{
P_{a+1,c-1,c,c+1}
P_{a,c,c+1,c+2}
}{
P_{c-1,c,c+1,c+2}
P_{a,a+1,c,c+1}
},
\]
with
\[
a=a_i,\qquad c=c_i,\qquad c-a=i+3.
\]
As $i=0,\ldots,n-6$, these are precisely the $n-5$ independent
constraints obtained in
Section~\ref{sec:$D{=}3$ kinematics limit}.
\end{proof}

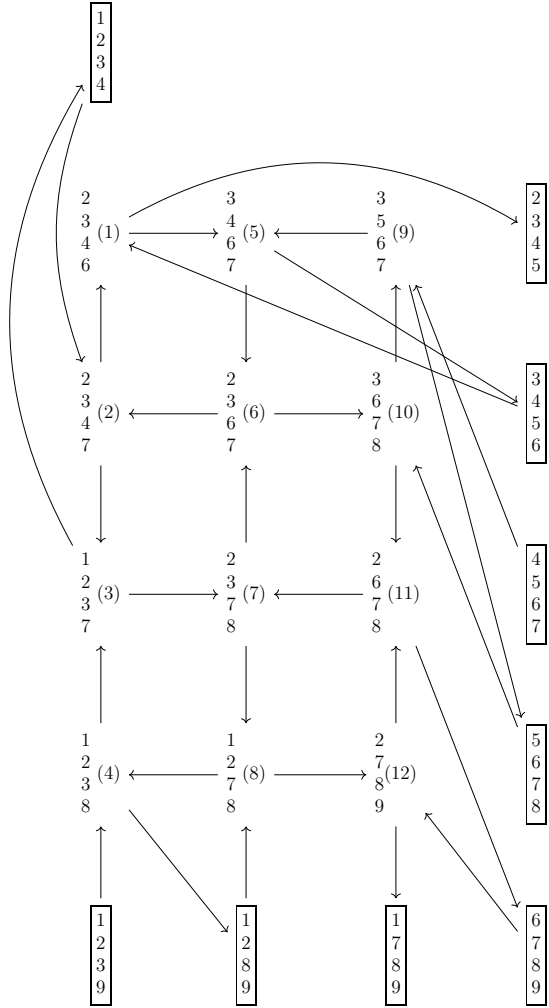
\begin{figure}
    \centering
    \adjustbox{scale=0.6,center}{%
\begin{tikzcd}
	{\fbox{$\begin{matrix} 1 \\ 2 \\ 3 \\4 \end{matrix}$}} \\
	\\
	{\begin{matrix} 2 \\ 3 \\ 4 \\6 \end{matrix} \ (1)} && {\begin{matrix} 3 \\ 4 \\ 6 \\7 \end{matrix} \ (5)} && {\begin{matrix} 3 \\ 5 \\ 6 \\7 \end{matrix} \ (9)} && {\fbox{$\begin{matrix} 2 \\ 3 \\ 4 \\5 \end{matrix}$}} \\
	\\
	{\begin{matrix} 2 \\ 3 \\ 4 \\7 \end{matrix} \ (2)} && {\begin{matrix} 2 \\ 3 \\ 6 \\7 \end{matrix} \ (6)} && {\begin{matrix} 3 \\ 6 \\ 7 \\8 \end{matrix} \ (10)} && {\fbox{$\begin{matrix} 3 \\ 4 \\ 5 \\6 \end{matrix}$}} \\
	\\
	{\begin{matrix} 1 \\ 2 \\ 3 \\7 \end{matrix} \ (3)} && {\begin{matrix} 2 \\ 3 \\ 7 \\8 \end{matrix} \ (7)} && {\begin{matrix} 2 \\ 6 \\ 7 \\8 \end{matrix} \ (11)} && {\fbox{$\begin{matrix} 4 \\ 5 \\ 6 \\7 \end{matrix}$}} \\
	\\
	{\begin{matrix} 1 \\ 2 \\ 3 \\8 \end{matrix} \ (4)} && {\begin{matrix} 1 \\ 2 \\ 7 \\8 \end{matrix} \ (8)} && {\begin{matrix} 2 \\ 7 \\ 8 \\9 \end{matrix} (12) } && {\fbox{$\begin{matrix} 5 \\ 6 \\ 7 \\8 \end{matrix}$}} \\
	\\
	{\fbox{$\begin{matrix} 1 \\ 2 \\ 3 \\9 \end{matrix}$}} && {\fbox{$\begin{matrix} 1 \\ 2 \\ 8 \\9 \end{matrix}$}} && {\fbox{$\begin{matrix} 1 \\ 7 \\ 8 \\9 \end{matrix}$}} && {\fbox{$\begin{matrix} 6 \\ 7 \\ 8 \\9 \end{matrix}$}}
	\arrow[from=11-1, to=9-1]
	\arrow[from=11-3, to=9-3]
	\arrow[from=9-1, to=11-3]
	\arrow[bend left=30, from=7-1, to=1-1]
	\arrow[bend left=-20, from=1-1, to=5-1]
	\arrow[from=3-1, to=3-3]
	\arrow[from=3-5, to=3-3]
	\arrow[from=3-3, to=5-3]
	\arrow[from=5-3, to=5-5]
	\arrow[from=5-5, to=3-5]
	\arrow[from=5-5, to=7-5]
	\arrow[from=7-5, to=7-3]
	\arrow[from=7-3, to=5-3]
	\arrow[from=5-3, to=5-1]
	\arrow[from=5-1, to=3-1]
	\arrow[from=5-1, to=7-1]
	\arrow[from=7-1, to=7-3]
	\arrow[from=7-3, to=9-3]
	\arrow[from=9-3, to=9-5]
	\arrow[from=9-5, to=7-5]
	\arrow[from=9-3, to=9-1]
	\arrow[from=9-1, to=7-1]
	\arrow[from=9-5, to=11-5]
	\arrow[from=11-7, to=9-5]
	\arrow[from=7-5, to=11-7]
	\arrow[from=9-7, to=5-5]
	\arrow[from=3-5, to=9-7]
	\arrow[from=7-7, to=3-5]
	\arrow[bend left=30, from=3-1, to=3-7]
	\arrow[from=5-7, to=3-1]
	\arrow[from=3-3, to=5-7]
\end{tikzcd}
}
\caption{The seed for $\Gr(4,9)$ after the mutation sequence $9,10,11,12,5,6,7,1,2,9,10,5$ starting from the initial seed in Figure \ref{fig:initial seed Gr49}.}
\label{fig:quiver for Gr49 after mutation sequence 9101112567910125}
\end{figure}

\begin{example}
\label{ex:Gr49 folding}
For $\Gr(4,9)$, the above construction gives the seed obtained from the initial seed by the mutation sequence
\[
9,10,11,12,5,6,7,1,2,9,10,5.
\]
The resulting seed is shown in Figure~\ref{fig:quiver for Gr49 after mutation sequence 9101112567910125}. The corresponding folding conditions are
\begin{align*}
& \frac{P_{3456}P_{2347}}{P_{2345}} = \frac{P_{4567}P_{3678}}{P_{5678}}, \quad \frac{P_{2346}P_{1237}}{P_{1234}} = \frac{P_{3567}P_{2678}}{P_{5678}}, \\
& \frac{P_{2347}P_{1238}}{P_{1234}} = \frac{P_{3678}P_{2789}}{P_{6789}}, \quad \frac{P_{1237}P_{1289}}{P_{1239}} = \frac{P_{1237}P_{9128}}{P_{9123}}  = \frac{P_{2678}P_{1789}}{P_{6789}}.
\end{align*} 
\end{example}

Computer program of mutations from the initial seed in Figure \ref{fig:initial seed Gr49} to the seed in Figure \ref{fig:quiver for Gr49 after mutation sequence 9101112567910125} can be found in \url{https://github.com/lijr07/folding-of-Grassmannian-cluster-algebras}. 

\begin{remark}
We note that [HLY26] provided evidence ($n \le 8$) for the relevance of folded cluster algebras to three-dimensional scattering amplitudes and Feynman integrals, in particular in ABJM theory. Our results establish a direct connection between three-dimensional kinematic constraints and foldings of Grassmannian
cluster algebras $\CC[\Gr(4, n)]$ for general $n$. This provides a systematic realization of the folding observed in \cite{HLY26}. It would be interesting to further investigate the physical implications of this structure, in particular, to what extent the resulting folded cluster algebras control the singularities or symbol
alphabets of scattering amplitudes and Feynman integrals in three-dimensional
theories.
\end{remark}

\end{document}